# Regime of a wideband phase-amplitude modulation in a CW magnetron transmitter with a phase control*

G. Kazakevich#, Northern Illinois University, Chicago, IL, USA,
R. Johnson, Muons, Inc., Batavia, IL, USA,
B. Chase, R. Pasquinelli, V. Yakovlev, Fermilab, Batavia, IL, USA,

A model of the CW high-power transmitter, utilizing frequency-locked magnetrons with a phase control studied initially as a prototype of controllable in phase and power an RF source for intensity-frontier superconducting linacs, was considered for telecommunication as a model of magnetron source, acceptable for a wideband phase-amplitude modulation at a precisely stable carrier frequency. The R&D conducted with CW, 2.45 GHz, 1 kW, microwave oven magnetrons demonstrated that the frequency locking of the magnetrons by the phase-modulated signal provides wideband phase and amplitude modulation at the modulating frequency at least up to 3 MHz and large magnitude, keeping the carrier frequency precisely stable, without broadening of the spectral line width. Performed experiments with power combining verified applicability of the transmitter based on the frequency-locked magnetrons for wideband phase and amplitude modulation, which may be used for telecommunication. Results of the experiments are described in the presented work.

## Introduction

Linear RF amplifiers as klystrons and IOTs are used traditionally in high-power transmitters prowiding power up to hundreds kW in CW mode at the carrier frequency in GHz range and the bandwidth of modulation in MHz range, that is acceptable for superconducting linacs and telecomunication. However, the cost of unit of power of the traditional RF sources is quite high, ~ \$5 per 1 W and ~ \$10 per 1 W for the RF sources based on klystrons and IOTs, respectively, [1, 2], since the capital costs of the high-power CW klystrons and IOTs and the respective environments costs are high. The CW magnetrons are less expensive in cost of unit of power and maintenance cost as well. For example, cost of unit of power of industrial, L-band, CW high-power magnetron RF source is ≈ \$1 per 1 W, [2]. Moreover the magnetron efficiency is highest in comparison with the traditional RF amplifiers. Thus the magnetron RF sources may be attractive candidates for some tasks in telecommunication. However, the traditional RF sources are amplifiers, providing frequency, phase and amplitude modulation at quite wide bandwidth, while the magnetrons are coherent auto-oscillators, frequency of which depends on the reflected signal and the magnetron current (frequency pulling and frequency pushing, respectively). Nevertheless, as it was demonstrated in [3-6], options of frequency and phase control (modulation) are inherent in magnetron-based transmitters in which, the magnetrons being quite well decoupled from the load, operate as forced (frequency-locked) oscillators. The wideband amplitude modulation first realized in model of the phase-controlled magnetron-based transmitter using the power combining and intended for superconductive linacs, also was experimentally verified, [5]. The wideband phase and amplitude modulation was provided by the transmitter model with the magnetrons frequency-locked by the phase-modulated signals.

The recent experiments with CW, S-band, 1 kW magnetrons, injection (frequency)-locked by the phase-modulated signal, [5, 6], demonstrated proof-of-principle of a wideband phase and amplitude control (modulation) in the magnetron transmitters. Description of the experiments and methods demonstrating the wideband phase and amplitude modulation in the magnetron transmitter model is presented and discussed in this work.

*Work was supported by the Fermilab and Muons, Inc. collaboration
#e-mail: gkazakevitch@yahoo.com

## Experimental technique and methods.

The experiments were performed with various setups using two magnetrons mounted in individual modules with necessary RF components, Fig. 1. The magnetrons with free run frequencies differing by 5.7 MHz operated being injection-locked at the frequency of 2.469 GHz.

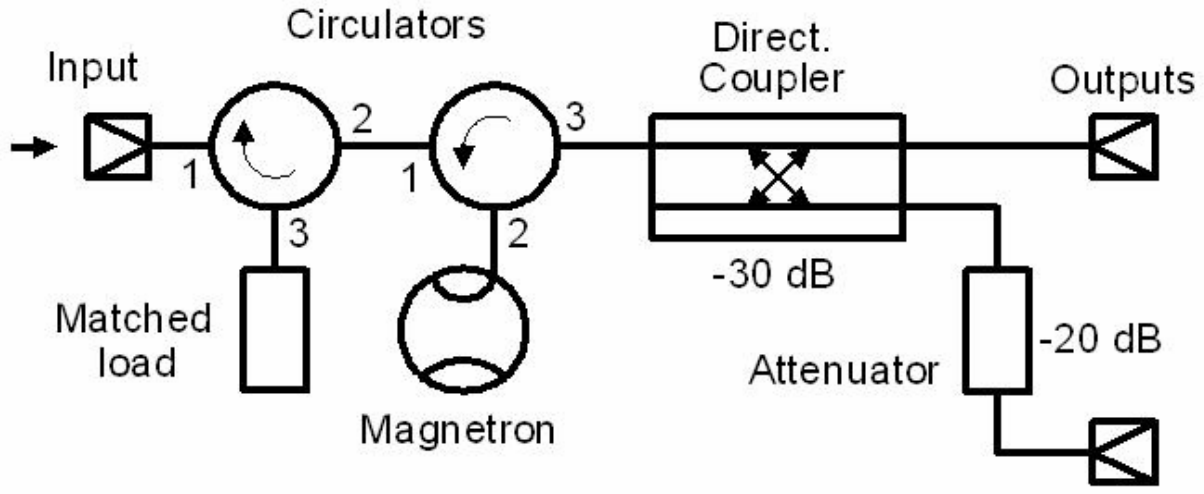

Fig. 1: The magnetron experimental module for tests of the transmitter setup.

The experiments were conducted generally in pulsed mode at large pulse duration (≈ 5 ms) accordingly requirements of R&D performed to study capabilities of magnetron transmitters allowing phase and power control for pulsed intensity-frontier superconducting linacs. Both magnetrons were fed by a single pulsed modulator with charging capacitor of 200 μF, providing droop of the modulator voltage ≤0.4 %. The magnetrons have dissimilar V-A characteristics; the magnetron with lower voltage was fed by a compensating divider.

Utilizing the modules in various setups we demonstrated proof-of-principle of wideband phase and amplitude modulation in the magnetron transmitter, Fig. 2, models based on the frequency-locked magnetrons, [7].

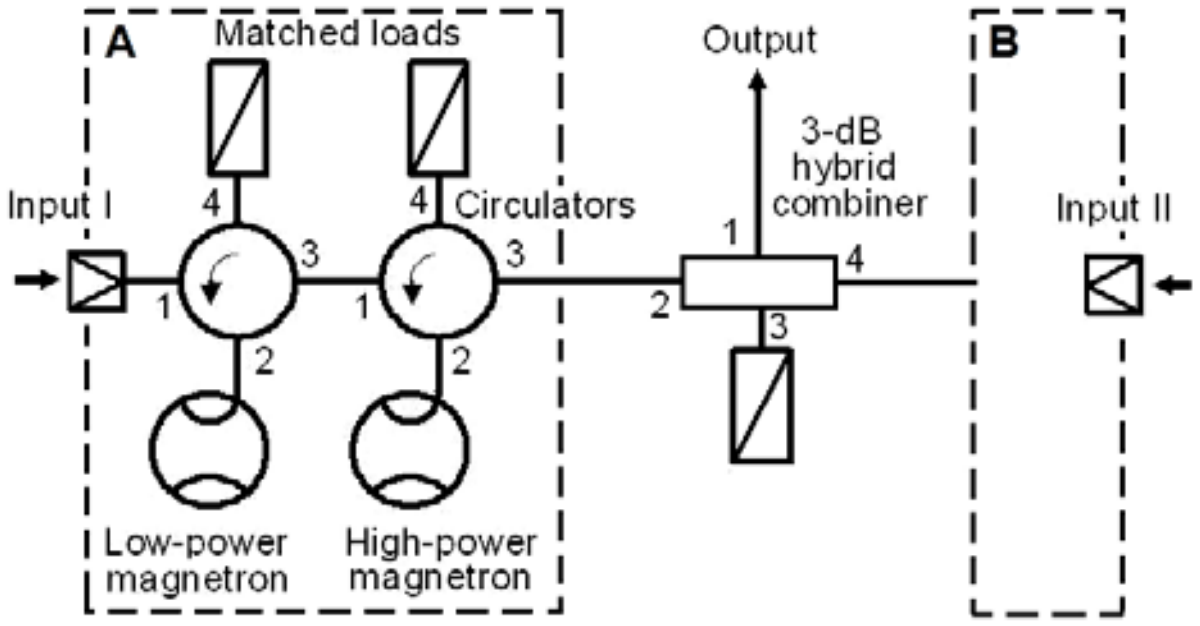

Fig. 2: Conceptual scheme of the magnetron transmitter acceptable for wideband phase and amplitude modulation.

The transmitter consists of two identical channels (A and B) of cascaded injection-locked magnetrons, combined in power by a 3-dB hybrid. The phase modulation is provided by modulation (simultaneously and equally) of the phases at inputs of both 2-cascade magnetrons. The amplitude (power) modulation is provided at modulation by phase difference at the inputs of the 2-cascade magnetrons. I.e. the amplitude modulation in the proposed magnetron transmitter is formally reduced to a phase modulation as well. The 2-cascade injection-locked magnetrons allow reducing the required locking power by 10-15 dB. More detailed description of the experimental technique is presented in [6].

Proof-of-principle of the wideband phase modulation in the transmitter was demonstrated in setup with single or 2-cascade magnetron, Fig. 3, frequency-locked by phase-modulated signal, [6].

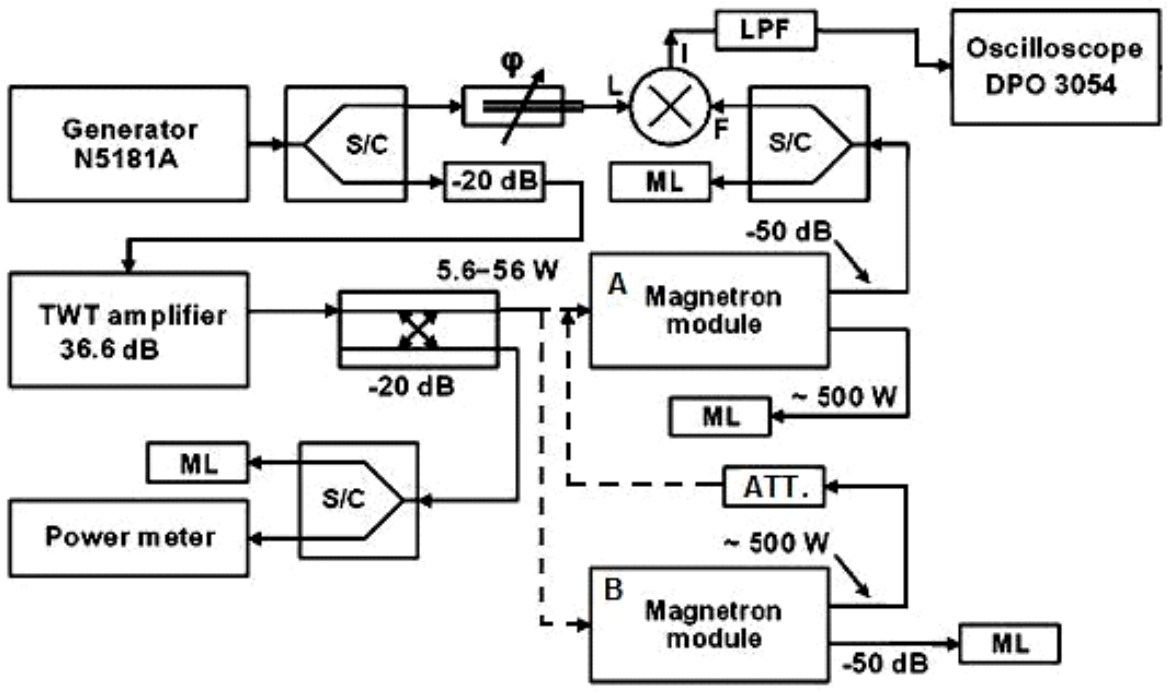

Fig. 3: Setup to study operation of single and 2-cascade frequency-locked magnetrons at the phase modulation. S/C is a splitter/combiner, LPF is a low pass filter, ML is a matched load, and ATT is an attenuator.

The single injection-locked magnetrons, Fig. 3, were tested in configuration using module A with the magnetron injection-locked by the CW TWT amplifier and fed by the modulator, while the module B was disconnected from the amplifier and the modulator. The 2-cascade magnetron was tested in configuration, in which the magnetron module B was frequency-locked by the TWT amplifier while the magnetron in module A was connected via attenuator to the module B output. Both magnetrons were fed by the modulator. In this case the magnetron in the module A was injection-locked by the pulsed signal of magnetron B, lowered in the attenuator. Experiments demonstrated operation of the 2-cascade magnetron in injection-locked mode at various values of the attenuator in the range of 13-20 dB, [6].

The wideband phase modulation of the single and 2-cascade frequency-locked magnetrons was realized at an internal phase modulation by a harmonic signal in the generator N5181A. The wideband (2-4 GHz) TWT amplifier did not distort the phase-modulated signal, frequency-locking the magnetrons.

The wideband phase and power (amplitude) modulation of the magnetrons combined in power were studied using setup shown in Fig. 4, [5, 6].

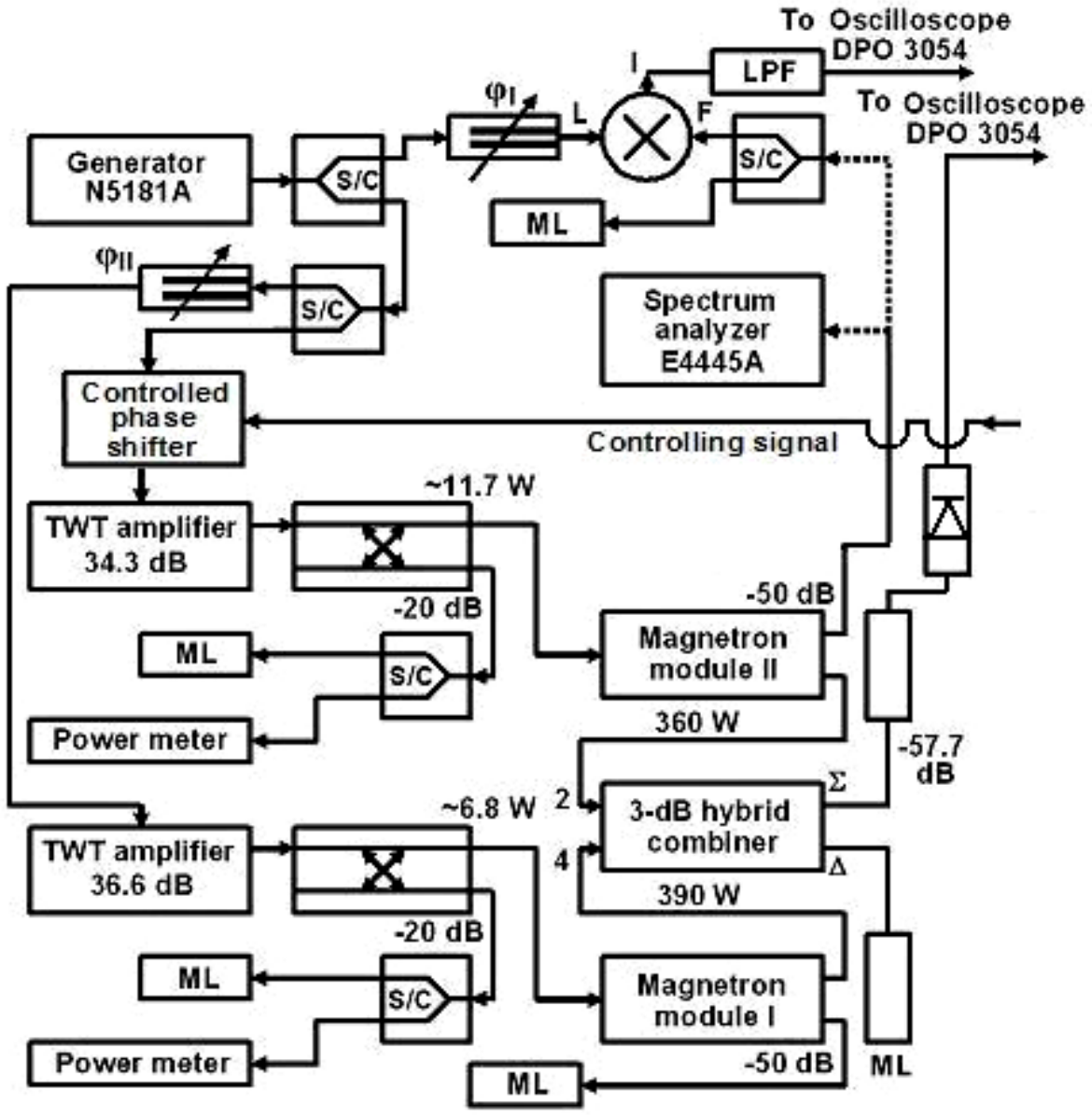

Fig. 4: Setup for the frequency-locked magnetrons with power combining to study wideband phase and power (amplitude) modulation in the transmitter.

Two kind of experiments were performed with this setup. At the first one was used the phase modulation by an internal harmonic signal in the generator at a constant phase shift in both channels, frequency-locking the magnetrons, combined in power by a 3-dB, 180 deg. hybrid. The trombone-like phase shifter $\varphi_{II}$ was tuned to get sum of magnetrons powers at the hybrid output "Σ". The controlled analogue phase shifter type JSPHS-2484 was OFF. In this experiment we measured the transfer function phase characteristic of the magnetrons with power summation at the phase modulation of the frequency-locking signal.

At the second experiment the phase modulation in the synthesizer was OFF, but the analogue phase shifter controlled by a sequence of rectangular pulses provided modulation of the phase difference of the frequency-locking signals in both magnetrons. This method first realized power (amplitude) modulation in the magnetron transmitter setup with power combining. Moreover measurement of the phase modulation at the output of the Magnetron module II was performed.

## Transient process in magnetrons frequency-locked by phase-modulated signal

We consider operation of the injection-locked magnetrons decoupled from the load. It is allowable considering high inverse losses of the state of the art ferrite circulators of the transmitter and an acceptable matching of the load impedance in the telecommunication systems. In this case one can neglect the frequency (phase) pulling in the magnetrons since the locking signal is much higher in amplitude than the reflected one.

As it was experimentally verified, [3, 4], variation of frequency (phase) of the signal, injection-locking the magnetron, causes a transient process, which is described with a good accuracy by an abridged equation, considering frequency pulling and frequency pushing in the magnetron. The abridged equation, [6] describing the transient process, caused by the phase pushing and modulation of the locking signal in the injection-locked magnetron, decoupled from the load, can be written as:

$$\left\{\frac{d}{dt}+\frac{\omega_{0M}}{2Q_{LM}}\left(1-i\varepsilon_M\right)\right\}\tilde{V}_M=\frac{\omega_{0M}}{Q_{EM}}\tilde{V}_{FM}-\frac{\omega_{0M}}{2Q_{EM}Y_{0M}}\tilde{I}_M. \quad (1)$$

Here: $Q_{LM}$, and $Q_{EM}$ are loaded and external magnetron Q-factors, respectively, $\omega_{0M}$ is the eigenfrequency of the magnetron cavity, $\varepsilon_M=\tan\psi\approx 2Q_{LM}\left(\omega_{0M}-\omega\right)/\omega_{0M}$ is the detuning parameter, $\omega$ is the frequency (time-dependent in common case) of the locking signal, $\tilde{V}_M$ and $\tilde{V}_{FM}$ are complex amplitudes of the oscillation in the magnetron cavity and in the wave locking the magnetron, respectively. $Y_{0M}=2\beta/R_{ShM}$ [1/Ohm] is the external waveguide conductance of the magnetron cavity, $\beta$ is the magnetron cavity coupling coefficient, $R_{ShM}$ is the magnetron cavity shunt impedance, $\tilde{I}_M$ is the complex amplitude of the first harmonic magnetron current. Terms $\omega_{0M}\cdot\tilde{V}_{FM}/Q_{EM}$ and $\omega_{0M}\cdot\tilde{I}_M/\left(2Q_{EM}\cdot Y_{0M}\right)$ describe modulation of the locking signal and phase pushing, respectively, $\psi$ is the angle between sum of the phasors $\tilde{V}_{FM}$ and $\tilde{I}_M$ and the phasor $\tilde{V}_M$, taken with corresponding coefficients, [8].

In steady-state the equation (1) is transformed into equation (2), accordingly to [8]:

$$\left|\tilde{V}_M\right|\cdot e^{i\psi}=\cos\psi\cdot e^{i\psi}\left(\frac{2Q_{LM}}{Q_{EM}}\cdot\left|\tilde{V}_{FM}\right|e^{i\psi}-\frac{Q_{LM}}{Q_{EM}\cdot Y_{0M}}\cdot\left|\tilde{I}_M\right|e^{i\psi}\right). \quad (2)$$

From this equation follows that variation of phase of the locking signal causes rotation of phasor of voltage in the magnetron cavity, Fig. 5, and in the wave at the magnetron output, $\tilde{V}_{MO}\approx\tilde{V}_M$, [6].

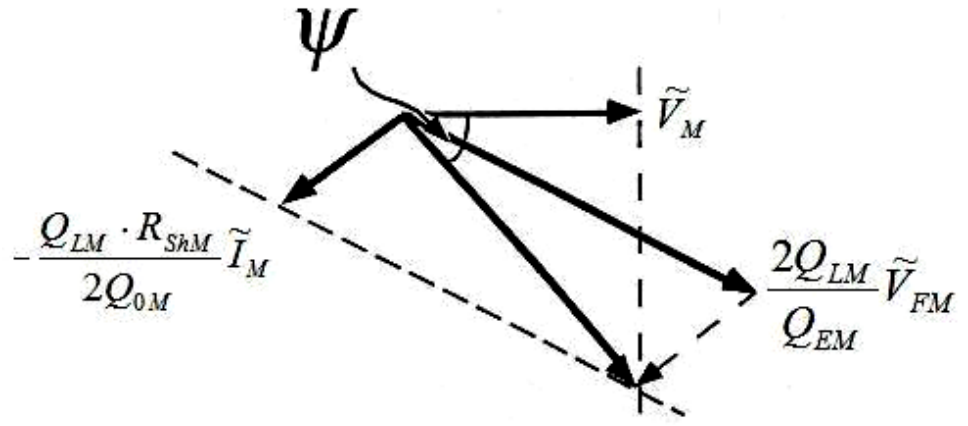

Fig. 5: Phasors diagram of decoupled from the load the injection-locked magnetron in steady-state.

The diagram substantiates the phase control (modulation) of magnetrons by phase-modulated locking signal. Assuming that the voltage in the magnetron cavity is approximately constant, for variations of locking signal and the magnetron current one can write the difference equation for imaginary terms, obtained from the equation (2):

$$\begin{aligned}&\left|\tilde{V}_M\right|\cos\psi\cdot\delta\psi\cong\\&-\sin\psi\cdot\sin 2\psi\cdot\left(\frac{2Q_{LM}}{Q_{EM}}\left|\tilde{V}_{FM}\right|-\frac{Q_{LM}}{Q_{EM}\cdot Y_{0M}}\left|\tilde{I}_M\right|\right)\delta\psi+\\&2\cos\psi\cdot\cos 2\psi\cdot\left(\frac{2Q_{LM}}{Q_{EM}}\left|\tilde{V}_{FM}\right|-\frac{Q_{LM}}{Q_{EM}\cdot Y_{0M}}\left|\tilde{I}_M\right|\right)\delta\psi+\\&\cos\psi\cdot\sin 2\psi\cdot\left(\frac{2Q_{LM}}{Q_{EM}}\delta\left|\tilde{V}_{FM}\right|-\frac{Q_{LM}}{Q_{EM}\cdot Y_{0M}}\delta\left|\tilde{I}_M\right|\right)\end{aligned}. \quad (3)$$

Since $Q_{LM}/Q_{EM}=\beta/\left(\beta+1\right)$, $\sin 2\psi=2\sin\psi\cdot\cos\psi$, $\cos 2\psi=\cos^2\psi-\sin^2\psi$, one gets:

$$\delta\psi\cong\frac{\sin 2\psi\left(\frac{2\beta}{\beta+1}\frac{\delta\left|\tilde{V}_{FM}\right|}{\left|\tilde{V}_M\right|}-\frac{R_{ShM}}{2(\beta+1)}\frac{\delta\left|\tilde{I}_M\right|}{\left|\tilde{V}_M\right|}\right)}{\left(6\sin^2\psi-2\right)\cdot\left(\frac{2\beta}{\beta+1}\frac{\left|\tilde{V}_{FM}\right|}{\left|\tilde{V}_M\right|}-\frac{R_{ShM}}{2(\beta+1)}\frac{\left|\tilde{I}_M\right|}{\left|\tilde{V}_M\right|}\right)+1}. \quad (4)$$

The equation (4) shows that variations of the magnetron current and/or the amplitude of the locking signal also causes rotation of the phasor of voltage in the magnetron cavity. Increase of the locking signal amplitude, $\left|\tilde{V}_{FM}\right|$, decreases the angle of rotation of the phasor $\tilde{V}_M$. Simplified analysis shows that the phase variation provides wider bandwidth of the phase control by a simple way.

## Experimental results

The transfer function phase characteristics in setups shown in Figs 3, 4 were measured by generator N5181A operating in the phase modulation mode, with harmonic modulating signal at magnitude of the phase modulation of 20 deg. The transfer function phase characteristics of the magnetrons were determined measuring angle $\theta$ of rotation of phasor of voltage, $\widetilde{V}_{MO} \approx \widetilde{V}_{M}$, in the magnetron output wave, relatively phasor of the frequency-locking signal, $\widetilde{V}_{FM}$, [6]. The phasor rotation results from a transient process caused by the phase modulation in the frequency-locked magnetron. The measurements were performed by phase detector, Figs. 3, 4, including phase shifter φ, ($\varphi_I$ in Fig. 4), double balanced mixer and Low Pass Filter (LPF).

The angle $\theta$ was determined as: $\theta \cong a\cos(1 - V_O / V_{PM})$, or $\theta \cong a\sin(V_O / V_{PM} - 1) + \pi/2$ at $V_O \le V_{PM}$ and $V_O > V_{PM}$, respectively, [6]. Here: $V_O$ is the voltage of the harmonic signal measured at output of the phase detector, $V_{PM}$ is the voltage, corresponding used magnitude of the phase modulation of 20 degrees. The angle $\theta$ of rotation of phasor $\widetilde{V}_{MO}$ vs. frequency of phase modulation, $f_{PM}$, at magnitude of the phase modulation of 20 degrees, for various setups and various values of locking signal power is plotted in Fig. 6.

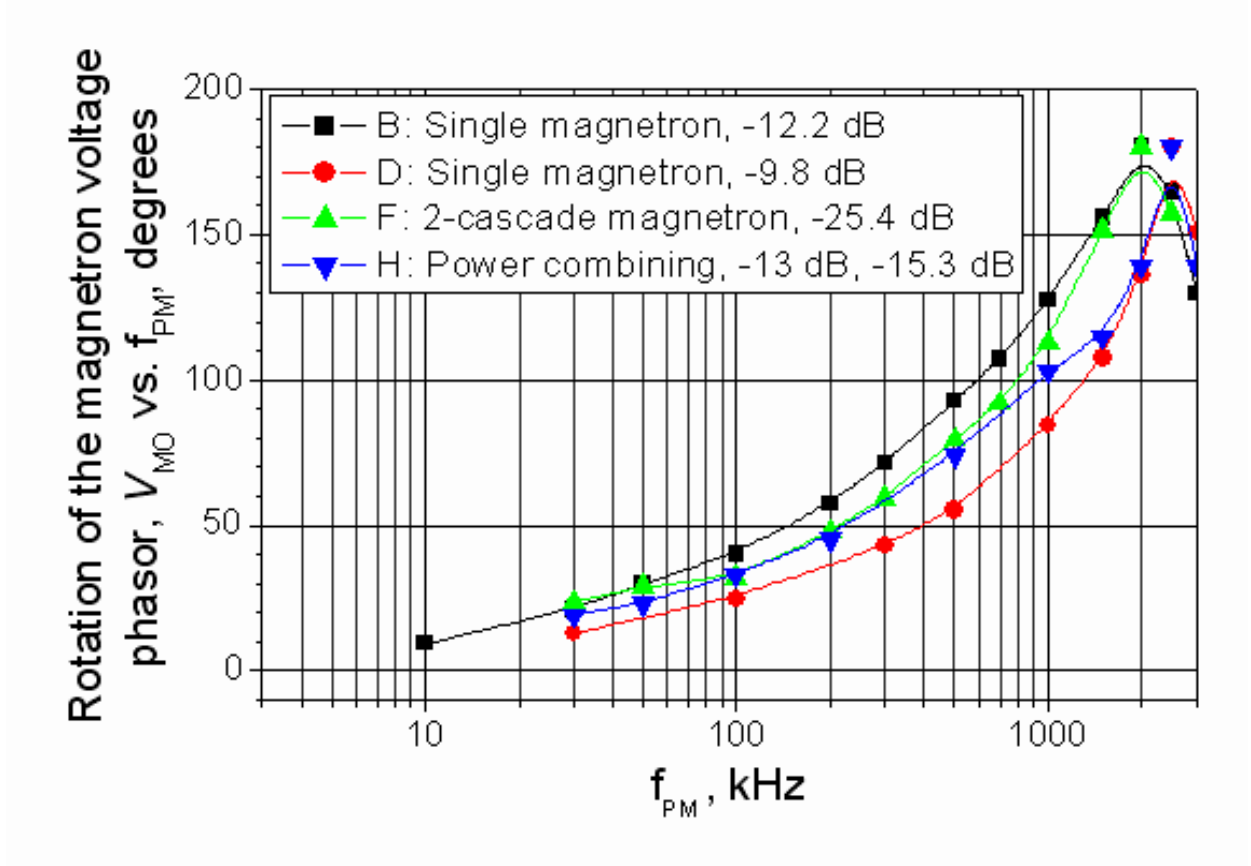

Fig. 6: Angle of rotation of the phasor of voltage in the wave at output of the frequency-locked magnetrons. The angle was measured vs. the modulating frequency, $f_{PM}$, in various setups, at various locking power.

At the measurements with power combining the controlled analogue phase shifter was OFF and the phase shifter $\varphi_{II}$ was tuned to get maximum power at the port "Σ" of the hybrid combiner.

From plots in Fig. 5 is seen that unlike amplifiers, rotation of phasor of the voltage in the magnetron output wave by 90 degrees does not disturb the carrier frequency stability of the injection-locked magnetron; the frequency-locked magnetron keeps the carrier frequency even the phasor rotates by ~ 180 degrees. Thus these values are not associated with pole on the phase characteristic. An increase of amplitude (power) of the frequency-locking signal decreases angle of rotation of the phasor at given magnitude of the phase modulation in accordance with equation (4). Note, that increase of the angle θ vs. $f_{PM}$ is approximately linear at $f_{PM} \geq 100$ kHz.

The phase delays, $-\theta/\omega_{PM} = -\theta/2\pi \cdot f_{PM}$, at various magnetron setups measured at phase modulation of the frequency-locking signal, Fig. 7, demonstrate quite smooth and flat plots in wide range (about decade) of modulating frequency $f_{PM}$. In this range the phase delay module is much less than 1 μs.

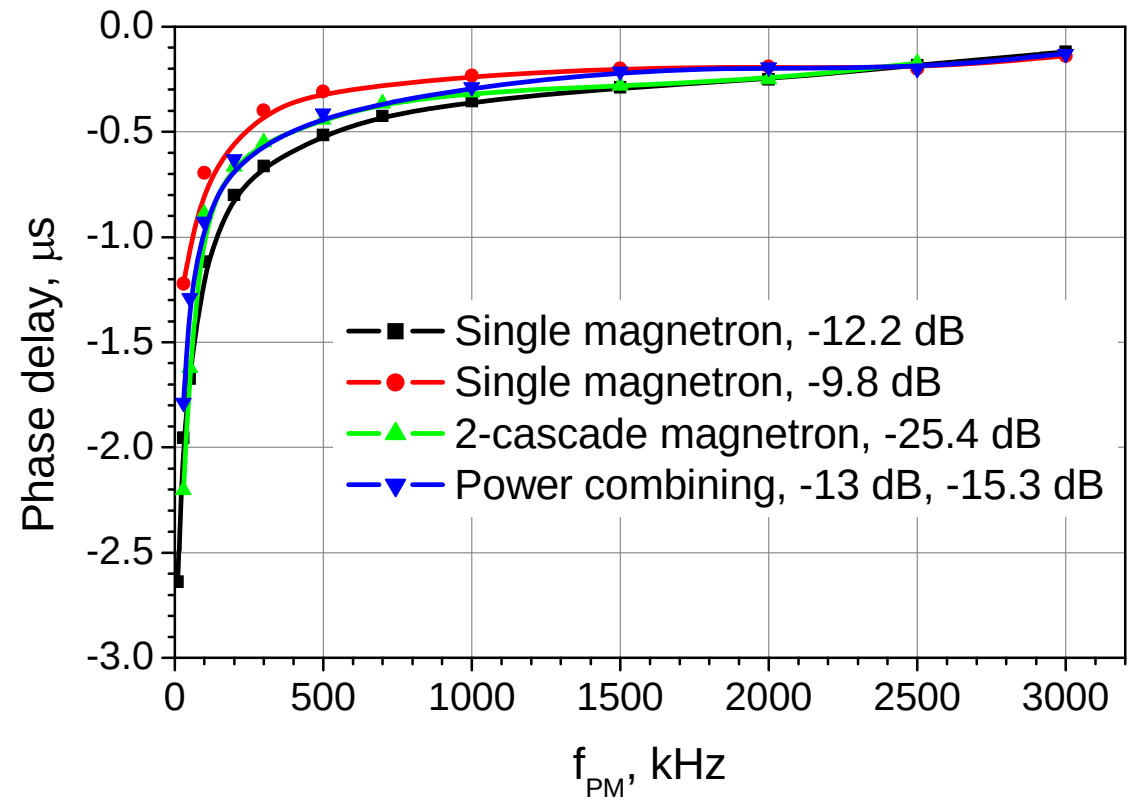

Fig. 7. Phase delays of the magnetrons frequency-locked by phase-modulated signal vs. $f_{PM}$.

The group delays of magnetrons, $\tau_g = d\theta/d\omega$, at the phase modulation of the frequency-locking signal, Fig. 8, have been determined by derivation of traces plotted in Fig. 7.

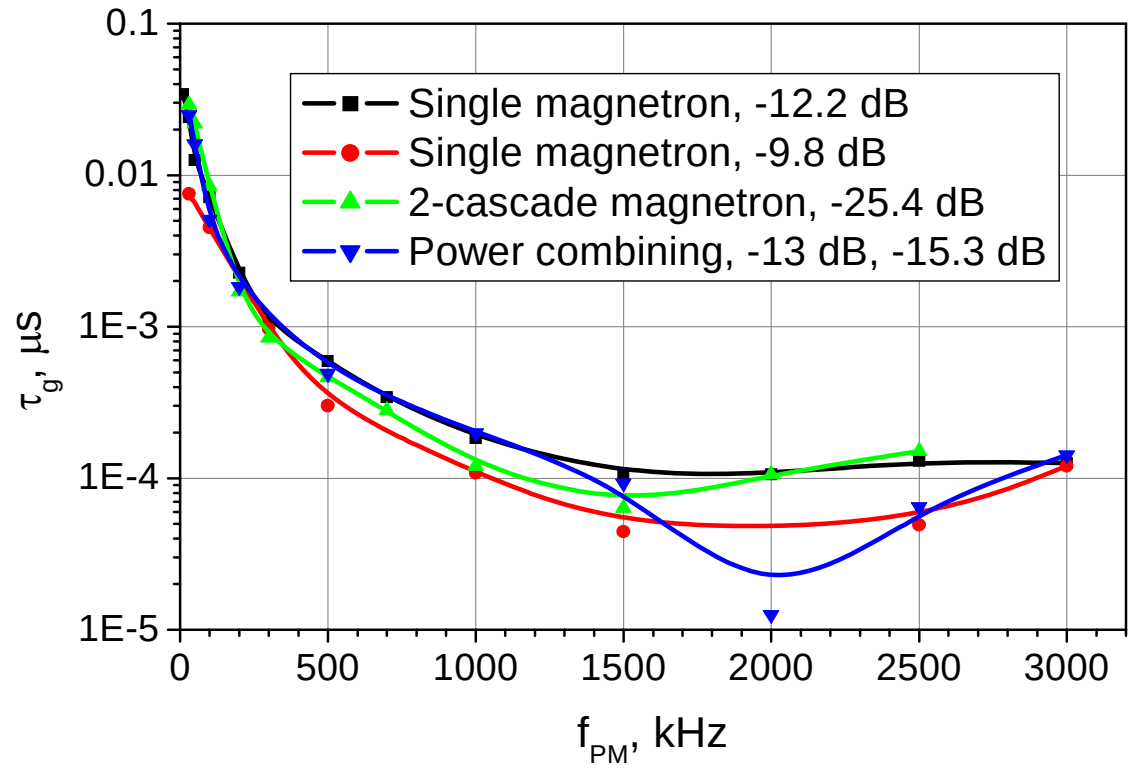

Fig. 8. Group delays of 2.45 MHz, CW, 1 kW magnetrons at various setups and at various values of power of the frequency-locking signal.

Traces of the Fig. 8 show that the group delays of the single, 2-cascade magnetron and magnetrons with power combining do not exceed 40 ns if the power of the locking signal (per magnetron in 2-cascade setup) is ~ -12 dB or more. The measured group delay at the phase modulation has a scale of the filling time of the magnetron cavity, $Q_{ML}/f_{M0}$, where $Q_{ML}$ is the loaded magnetron Q-factor, $f_{M0}$ is the carrier frequency of the magnetron. Increase of the locking signal decreases the group delay of the frequency-locked magnetrons. One can understand this, since the transient processes in magnetrons, lasting during many periods of oscillations, are averaged over the filling time which is ~ of the group delay. Thus one can expect, that the carrier frequency of the magnetron, frequency-locked by the phase-modulated signal, should be stable if $\theta < 2\pi$ during the filing time interval. Note that an increase of amplitude of the of the frequency-locking signal $|\widetilde{V}_{FM}|$, decreases angle of rotation of the phasor $\widetilde{V}_M$, decreasing the group delay and the averaged angle caused by the phase modulation during the time interval.

The transfer function magnitude characteristics also indicate capabilities of the wideband phase modulation in the magnetron transmitter. The transfer function magnitude characteristics at the phase modulation of the frequency-locking signal, Fig. 9, were measured for single and 2-cascade magnetrons at low magnitude (70 mrad. ≈ 4 deg.) of the modulating signal. The measurements were performed by the Agilent MXA N9020A Signal Analyzer in the phase modulation domain for various ratios the magnetrons power to power of the locking signal.

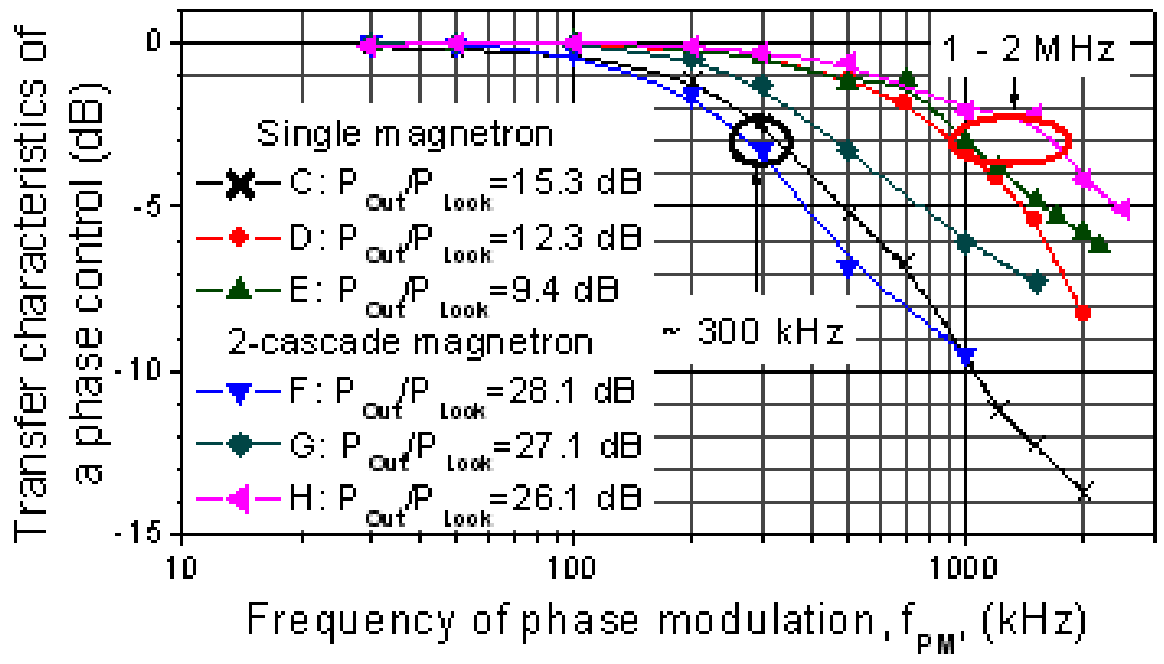

Fig. 9. Transfer function magnitude characteristics (rms values) at the phase modulation measured in phase modulation domain with single and 2-cascade injection-locked magnetrons.

Plotted in Fig. 9 the transfer function magnitude characteristics (rms values) were averaged over 8 pulses for the injection-locked single 2M219J magnetron and the 2-cascade magnetron setup. Non-flatness of the phase characteristic of generator N5181A has been measured and taken into account, [6].

The plots demonstrate 3-dB cutoff of the magnitude characteristics for single and 2-cascade magnetron over of 1 MHz at the locking power of more than -12.3 dB per magnetron. Increase of the locking power increases the frequency cutoff of the phase modulation.

Plots in Figs. 5-9 demonstrate capabilities of the frequency-locked magnetron transmitter for wideband phase modulation. Fig. 10 demonstrates wideband amplitude (power) modulation first performed by magnetrons with power combining, using setup shown in Fig. 4.

The amplitude modulation was realized controlling the analogue phase shifter JSPHS-2484 by sequence of rectangular pulses of voltage with period of ≈30 μs. At the measurements the phase shifter $\varphi_{II}$ was tuned to get maximum power at the port "Σ" when the signal controlling the phase shifter JSPHS-2484 was OFF. The phase detector trombone $\varphi_{I}$ was tuned to avoid saturation at measurements of the phase modulation of the magnetron in the Magnetron module II. Note that limitation of bandwidth seen in plots Fig. 10, [6], results from limited bandwidth (~ 50 kHz) of the used analogue phase shifter at the voltage control.

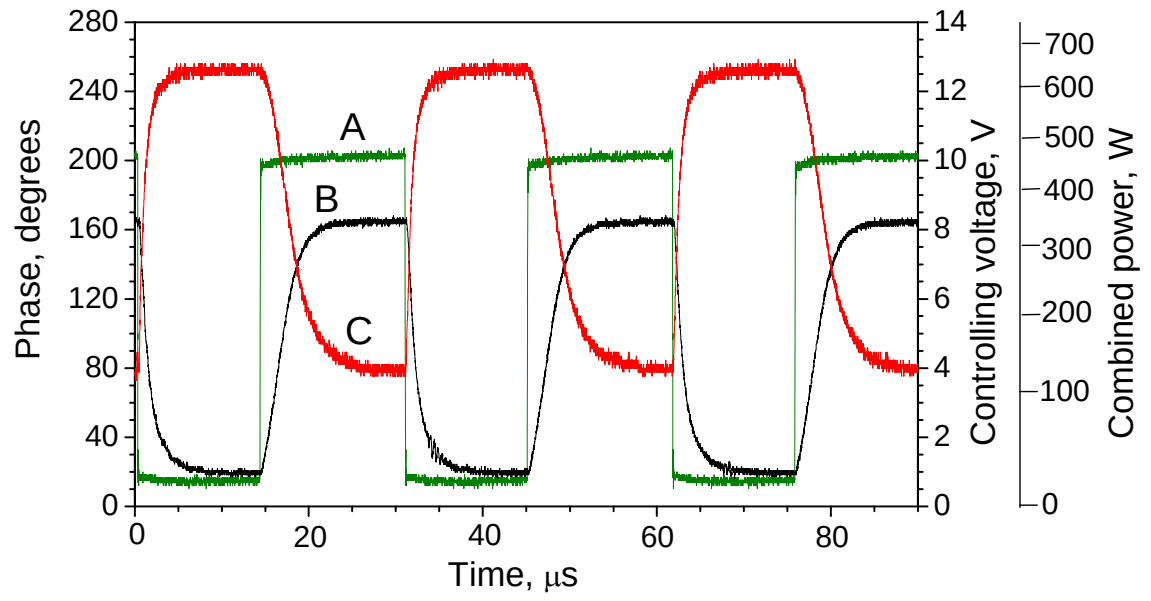

Fig. 10: Amplitude (power) modulation of the frequency-locked magnetrons with power combining. Trace A shows shape of signals controlling the phase shifter, first right scale. Trace B is the phase variations at the output of the magnetron II measured by the phase detector, left scale. Trace C shows power measured by a calibrated detector at port "Σ" of the hybrid combiner at the phase shifter pulsed voltage control, second right scale.

Plots in Figs. 5-10 demonstrate proof-of-principle of wideband phase and amplitude modulation performed with model of the high-power transmitter based on the frequency-locked magnetrons.

Fig. 11, [6], shows that the carrier frequency of the transmitter with highest accuracy repeats the carrier frequency of the locking signal at wide range of the modulating frequency and magnitude.

The measurements, [6], at the resolution bandwidth of 1.0 Hz were performed by the Agilent MXA N9020A Signal Analyzer using a 2.45 GHz, 1 kW magnetron, operating in CW mode at locking power of -13.4 dB, output power of ~ 850 W, carrier frequency of the locking signal of 2.451502 GHz. Setup of the measurements looks like shown in Fig. 1. Loss of power at the carrier frequency at large magnitude of the phase modulation, Fig. 10, trace B, results from redistribution of the magnetron power into sidebands at large index of the modulation. The sidebands differing from the carrier frequency by 11.3 and 15 Hz result from the locking system.

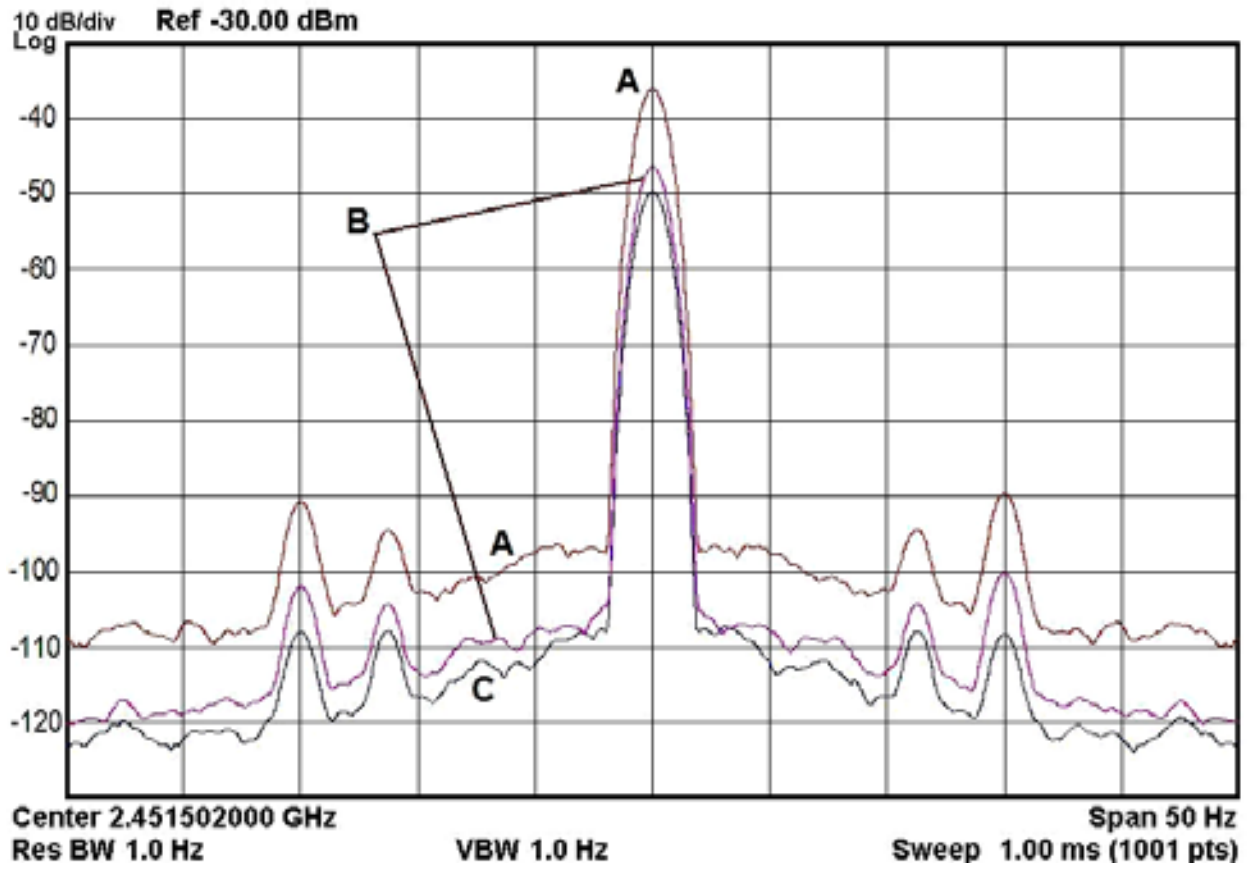

Fig. 11: Carrier frequency spectra of the magnetron injection-locked by a signal without, trace A, and with the phase modulation, trace B. Trace B shows the spectrum of the magnetron injection-locked by the phase-modulated signal at magnitude and frequency of the modulation of 3 radians and 2 MHz, respectively. Trace C shows the carrier frequency spectrum of the locking system (N5181A generator and TWT amplifier) without the phase modulation. Scales in vertical and horizontal are: 10 dB/div and 5 Hz/div, respectively.

No broadening of the carrier frequency spectra was observed in the range of modulating frequency of (0-3 MHz) and large magnitude (up to few radians) of the phase modulation. Plotted in Fig. 10 the magnetron carrier frequency spectrum width at the level of -60 dBc is 3.8 Hz, while the spectrum width of the locking signal is 3.68 Hz at the same level. This indicates that the own carrier frequency spectrum width of the injection-locked magnetron, measured at the level of -60 dBc at the resolution bandwidth of 1.0 Hz is ~1 Hz.

Note, that first experiments with 2.5 MW, S-band, pulsed magnetron injection-locked by frequency-varied (modulated) locking signal, [3, 4], verified capabilities of the magnetrons for frequency modulation, however in accordance with requirements of feeding of superconducting linacs, for which the R&D with 2.45 GHz, 1 kW magnetrons was conducted, we considered here the options of the phase and amplitude modulation only. Obtained results at the phase modulation allow to expect that capabilities of the injection-locked magnetrons in bandwidth at the frequency modulation should be approximately same.

The transition processes in the frequency-locked magnetrons, caused by ripple in magnetron current or/and disturbances of the magnetic field induced by magnetron filament circuitry, result in phase pushing, [5, 6], which may distort the phase modulation, however the low-frequency phase distortions can be suppressed in demodulators by appropriate high pass filters.

The nonlinear distortions in the magnetron transmitter, caused by dependence of $\theta(f_{PM})$ at the phase modulation of the injection (frequency)-locking signal can be considered by an appropriate digital modulation-demodulation system.

On can hope that the efficient and lower cost magnetron CW transmitters can be used in telecommunication systems utilizing wideband amplitude, phase, frequency and/or Quadrature Amplitude Modulation (QAM).

## Summary

The transmitter, proposed for superconducting intensity-frontier linacs, based on frequency-locked magnetrons is acceptable for a wideband phase and amplitude modulation at the precisely-stable carrier frequency. Experiments conducted with CW, 2.45 GHz, 1 kW magnetrons frequency-locked by the wideband phase-modulated signal demonstrated Proof-of-Principle of the wideband phase and amplitude modulation with bandwidth up to 3 MHz in models of the injection-locked magnetron transmitter.